# Coronavirus and financial volatility: 40 days of fasting and fear

Claudiu Tiberiu ALBULESCU[1,2*]

[1] Management Department, Politehnica University of Timisoara, 2, P-ta. Victoriei, 300006 Timisoara, Romania.

[2] CRIEF, University of Poitiers, 2, Rue Jean Carbonnier, Bât. A1 (BP 623), 86022, Poitiers, France.

**Abstract**

40 days after the start of the international monitoring of COVID-19, we search for the effect of official announcements regarding new cases of infection and death ratio on the financial markets volatility index (VIX). Whereas the new cases reported in China and outside China have a mixed effect on financial volatility, the death ratio positively influences VIX, that outside China triggering a more important impact. In addition, the higher the number of affected countries, the higher the financial volatility is.

---

[*] E-mail: claudiu.albulescu@upt.ro.

# 1. Introduction

The coronavirus (COVID-19) outbreaks in December 2019 in China, in the city of Wuhan (Hubei region). On March 3, 2020, the virus has already affected more than 90,000 people in more than 60 countries, having killed thousands. Starting with January 20, 2020, the World Health Organization (WHO) monitored the situation and released daily reports about new cases of infections and death number in the Chinese regions and outside China. Following the general fear in China, the Shanghai stock market plunged 8% on February 3, 2020, and the shock rapidly spread over international financial markets. The United States (US) stock prices recorded their lower level in the last six months. However, a second, and more important shock hit the US market on February 28, when the S&P 500 plummeted 4.4%. Initially ignored, the COVID-19 effect raised serious concerns since the infection rapidly propagated outside China. The WHO report of February 28, 2020 underlined over one thousand new cases outside China and five new countries affected. The US financial markets reacted to this news, although Mr. Trump announced that a solution to this problem will be found "soon", by the scientists.

The coronavirus panic affects the world economy, with a negative impact on trade and tourism, generating local food shortages.[1] In addition, in the presence of stock markets price bubbles (S&P 500 recorded a maximum of 3,380 points on February 14, 2020, meaning an increase of 65%, as compared to February 14, 2015), COVID-19's impact on the financial system cannot be ignored. Therefore, several questions emerge. How does this virus affect financial markets volatility? Will COVID-19 be the source of a new financial crisis? Without tempting to provide a straight answer to the second question, the purpose of this paper is to show how the coronavirus figures reported by WHO impact the financial markets volatility index (VIX). We consider a period of 40 days, starting with January 20, 2020, up to February 28, 2020. We look at three categories of data, namely new case announcements and death ratio (in China and outside China), as well at the number of daily affected countries.

The financial volatility has different sources, related to economic conditions, institutional issues or market uncertainty (Hartwell, 2018). Macroeconomic announcements also affect the financial volatility. In this line, Onan et al. (2014) find that good and bad announcements asymmetrically impact VIX, whereas most of recent studies focus on the role of Economic Policy Uncertainty (EPU) in influencing the financial volatility (Antonakakis et al., 2013; Chen and Chiang, 2020; Kalyvas et al., 2019; Mei et al., 2018; Tiwari et al., 2019; Zhenghui and

---

[1] In Northern Italy, several supermarkets were emptied, situation also recorded in other European countries, as Germany and the United Kingdom. Consequently, the "fasting" period installed quite earlier this year.

Junhao, 2019). For example, Karnizova and (Chris) Li (2014) predict the US recession using the interaction between EPU and stock market volatility, whereas Zhu et al. (2019) investigate how a fear index influences the US stock market volatility.

Looking to the coronavirus-generated fear, this is the first paper addressing the impact of COVID-19 announcements on financial market volatility. We discover that the figures related to the propagation of the coronavirus outside China generate a higher volatility on financial markets, as compared to those reported for China. Further, the financial volatility increases with the number of affected countries.

## 2. Coronavirus: stylized facts

COVID-19 started to be considered more aggressive as compared to the Severe Acute Respiratory Syndrome (SARS) recorded in 2003. The number of deaths already overlaps those generated by SARS, spreading in more than 60 countries at the beginning of March 2020. According to WHO data, the virus regresses in China, but rapidly spreads in countries like South Korea, Italy or Iran. Figure 1 presents the dynamics of the total number of infected persons, new cases, death ratio and affected countries since January 20, 2020.

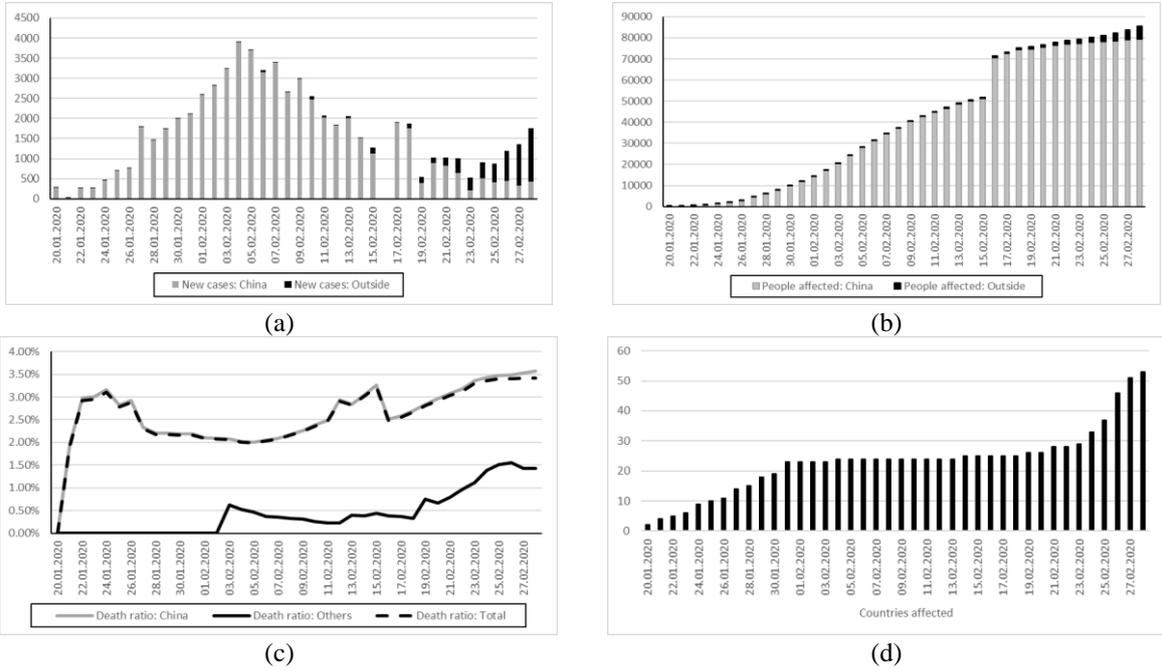

Fig. 1. COVID-19 dynamics
*Source: WHO situation reports*

According to Figure 1(a) the pick of new cases announced in China was reached in February. In fact, on February 16 (the associated statistics are intentionally excluded from the graphic), China announced a record of 19,461 new cases. Figure 1(b) shows that most of infected people originate in China. The death ratio computed as the ratio between reported deaths and infected persons (Figure 1(c)) is much lower outside China, as compared to China. However, the overall death ratio is close to 3.5%. Finally, since February 23, we have witnessed a rapid spread of coronavirus at global level (Figure 1(d)).

**3. Empirical specification and results**

We test a simple regression on the coronavirus effect on financial volatility and we use a stepwise procedure. In the first step (Eq. (1)) we implement a naïve estimation whereas in the second step (Eq. (2)) we consider the US EPU as control variable. Daily data comes from WHO and FRED database, respectively.

$$VIX_t = c + \alpha_t COVID - 19_t + \varepsilon_t \quad (1)$$

$$VIX_t = c + \alpha_t COVID - 19_t + \beta_t EPU_t + \varepsilon_t \quad (2)$$

We estimate three models, with a focus on China's reported data (Model 1), on countries outside China (Model 2) and on the overall situation (Model 3). Three types of analyzes are performed. In Table 1 we present the impact of announcements related to new infection cases.[2] In Table 2 we show the estimates of death ratio influence on financial volatility. Finally, in Table 3 we show the results of the impact of coronavirus worldwide spread.

Table 1 highlights that COVID-19 new reported cases have a marginal negative effect on financial volatility (Model 1). This effect remains unchanged if EPU is used as control variable and can be explained by a decrease of the number of new infections in China in the second part of February, while the volatility slightly increased. Nevertheless, if we look to the new cases reported outside China (Model 2), we clearly see that COVID-19 contributes to an increase of financial volatility. Likewise, the markets are more sensitive to the coronavirus spillover in Europe and US. Model 3 reveals inconclusive results about the COVID-19 total new reported cases on VIX.

---

[2] For this set of analyses and Model 1 (where we investigate the impact of new cases reported in China on VIX), we also apply an ordinary least square regression with a structural break, to account for the potential effect of new cases reported by China on February 16, 2020. The results are not significantly different from those shown in Table 1 and can be provided under request.

Table 1. New case announcements and financial volatility

| New cases | Model 1 – China | | Model 2 – Outside | | Model 3 – Total | |
| --- | --- | --- | --- | --- | --- | --- |
| | Naïve | Control | Naïve | Control | Naïve | Control |
| COVID-19 | -0.002* | -0.002** | 0.020*** | 0.020*** | -0.000 | -0.001 |
| | [0.001] | [0.000] | [0.001] | [0.001] | [0.001] | [0.001] |
| EPU | | 0.111*** | | 0.014 | | 0.113*** |
| | | [0.034] | | [0.013] | | [0.038] |
| c | 21.54*** | 11.32*** | 14.53*** | 13.22*** | 19.15*** | 9.819** |
| | [2.198] | [3.659] | [0.460] | [1.298] | [2.676] | [3.957] |
| $R^2$ | 0.116 | 0.379 | 0.915 | 0.919 | 0.005 | 0.260 |

*Notes: (i) 10%, 5% and 1% level of significance is denoted by \*, \*\* and \*\*\* respectively; (ii) COVID-19 is associated with the new reported cases.*

In a subsequent analysis we investigate the impact of COVID-19 death ratio on VIX (Table 2). The death ratio has a positive and very significant impact on financial volatility and this result is very robust. Moreover, as expected, we notice that the death ratio associated with the virus propagation outside China determines a higher financial volatility, as compared to the ratio reported for China. In terms of elasticities, the naïve specification (Model 2) evidences that an increase of 1% in the death ratio generates an increase of 11% in the financial volatility index. Compared to the previous set of estimations, where no significant effect is reported, the total death ratio (China and other affected countries) positively impacts the volatility.

Table 2. Coronavirus death ratio and financial volatility

| New cases | Model 1 – China | | Model 2 – Outside | | Model 3 – Total | |
| --- | --- | --- | --- | --- | --- | --- |
| | Naïve | Control | Naïve | Control | Naïve | Control |
| COVID-19 | 7.840*** | 7.629*** | 11.07*** | 10.10*** | 7.847*** | 7.750*** |
| | [2.081] | [1.771] | [1.606] | [1.696] | [2.234] | [1.901] |
| EPU | | 0.095*** | | 0.039 | | 0.097*** |
| | | [0.028] | | [0.026] | | [0.029] |
| c | -2.798 | -11.54** | 12.87*** | 9.449*** | -2.572 | -11.76* |
| | [5.702] | [5.526] | [1.137] | [2.528] | [6.022] | [5.831] |
| $R^2$ | 0.353 | 0.549 | 0.646 | 0.675 | 0.323 | 0.528 |

*Notes: (i) 10%, 5% and 1% level of significance is denoted by \*, \*\* and \*\*\* respectively; (ii) COVID-19 is associated with the death ratio.*

The last analysis tests the effect of the number of countries affected by the virus during the analyzed period, on the financial volatility. Table 3 shows a positive impact on VIX, confirming thus the previous results. The findings are robust to the use of US EPU as a control variable and show that the spread of coronavirus in general, and the increasing death ratio in particular, bust the financial volatility.

Table 3. Affected number of countries and financial volatility

| Countries | Naïve | Control |
|---|---|---|
| COVID-19 | 0.506*** | 0.487*** |
|  | [0.063] | [0.073] |
| EPU |  | 0.013 |
|  |  | [0.026] |
| c | 6.009*** | 5.134** |
|  | [1.706] | [2.426] |
| $R^2$ | 0.710 | 0.713 |

*Notes: (i) 10%, 5% and 1% level of significance is denoted by \*, \*\* and \*\*\* respectively; (ii) COVID-19 is associated with the affected number of countries.*

## 4. Conclusions

We have tested the impact of COVID-19 official announcements and related figures on the financial volatility, comparing the effect of data reported in China, with that of COVID-19 numbers reported outside China. Our results show that: (i) only the new cases reported outside China have a positive impact on VIX, (ii) the death ratio has a significant and positive impact on VIX for all tested models, and the effect is stronger for the death ratio outside China, (iii) the spread of coronavirus increases the financial volatility. The persistence of COVID-19 might generate a new episode of international financial stress.